\documentclass[
 aps,
reprint,%
superscriptaddress,
]{revtex4-1}
\usepackage{graphicx}
\usepackage{dcolumn}
\usepackage{bm}
\usepackage{verbatim}
\usepackage{braket}
\usepackage{enumerate}
\usepackage{tabularx}
\usepackage{xcolor}
\usepackage{dcolumn}

\begin{document}


\title{Demonstration of Laser-produced Neutron  Diagnostic  by Radiative Capture Gamma-rays}

\def\INPAC{INPAC and School of Physics and Astronomy, Shanghai Jiao Tong University, Shanghai 200240, China}
\def\shKeyLab{Shanghai Key Laboratory for Particle Physics and Cosmology, Shanghai Jiao Tong University, Shanghai 200240, China}
\def\LLP{Key Laboratory for Laser Plasmas (MoE), School of Physics and Astronomy, Shanghai Jiao Tong University, Shanghai 200240, China}
\def\IFSA{Collaborative Innovation Center of IFSA (CICIFSA), Shanghai Jiao Tong University, Shanghai 200240, China}
\def\IOP{Beijing National Laboratory for Condensed Matter Physics, Institute of Physics, Chinese Academy of Sciences, Beijing 100190, China}
\def\UCAS{School of Physical Sciences, University of Chinese Academy of Sciences, Beijing 100049, China}
\def\USTC{University of Science and Technology of China, Hefei 230026, China}
\def\SIOM{Shanghai Institute of Optical and Fine Mechanics, Chinese Academy of Sciences, Shanghai 201800, China}
\def\CIAE{Department of Nuclear Physics, China Institute of Atomic Energy, Beijing 102413, China}
\def\SILP{Shanghai Institute of Laser Plasma, China Academy of Engineering Physics, Shanghai 201800, China}
\def\SINAP{Shanghai Institute of Applied Physics, Chinese Academy of Sciences, Shanghai 201800, China}
\def\SUPA{SUPA, Department of Physics, University of Strathclyde, Glasgow G4 0NG, UK}

\author{Xiaopeng Zhang}
\address{\INPAC}\address{\shKeyLab}
\author{Wenqing Wei} \address{\LLP}\address{\IFSA}
\author{Changbo Fu}
\email[Corresponding author: ]{cbfu@sjtu.edu.cn}
\address{\INPAC}\address{\shKeyLab}
\author{Xiaohui Yuan}
\email[Corresponding author: ]{xiaohui.yuan@sjtu.edu.cn}
\address{\LLP}\address{\IFSA}
\author{Honghai An} \address{\SILP}
\author{Yanqing Deng} \address{\LLP}\address{\IFSA}
\author{Yuan Fang} \address{\LLP}\address{\IFSA}
\author{Jian Gao} \address{\LLP}\address{\IFSA}
\author{Xulei Ge} \address{\LLP}\address{\IFSA}
\author{Bing Guo} \address{\CIAE}
\author{Chuangye He} \address{\CIAE}
\author{Peng Hu} \address{\USTC}
\author{Neng Hua} \address{\SIOM}
\author{Weiman Jiang} \address{\IOP}
\author{Liang Li} \address{\INPAC}\address{\shKeyLab}
\author{Mengting Li} \address{\USTC}
\author{Yifei Li} \address{\IOP}
\author{Yutong Li} \address{\IFSA}\address{\IOP}\address{\UCAS}
\author{Guoqian Liao} \address{\LLP}\address{\IFSA}
\author{Feng Liu} \address{\LLP}\address{\IFSA}
\author{Longxiang Liu} \address{\SINAP}
\author{Hongwei Wang} \address{\SINAP}
\author{Pengqian Yang} \address{\SIOM}
\author{Su Yang} \address{\LLP}\address{\IFSA}
\author{Tao Yang} \address{\USTC}
\author{Guoqiang Zhang} \address{\SINAP}
\author{Yue Zhang} \address{\SINAP}
\author{Baoqiang Zhu} \address{\SIOM}
\author{Xiaofeng Xi} \address{\CIAE}
\author{Jianqiang Zhu} \address{\SIOM}
\author{Zhengming Sheng} \address{\LLP}\address{\IFSA}\address{\SUPA}
\author{Jie Zhang} \address{\LLP}\address{\IFSA}

\date{\today}

\begin{abstract}
We report a new scenario of time-of-flight (TOF) technique in which fast neutrons and delayed gamma-ray signals were both recorded in a millisecond time window in harsh environments induced by high-intensity lasers.
The delayed gamma signals, arriving far later than the original fast neutron and often being ignored previously,
were identified to be the results of  radiative captures of thermalized neutrons.
The linear correlation between gamma photon number and the fast neutron yield shows that these delayed gamma events can be employed for neutron diagnosis. 
This method can reduce the detecting efficiency dropping problem caused by prompt high-flux gamma radiation, 
and provides a new way for neutron diagnosing in high-intensity laser-target interaction experiments.
\end{abstract}
\maketitle


\section{Introduction}

There has been a growing interest in laser-driven pulsed neutron sources, for their wide applications in fields like fundamental physics~\cite{kappeler_nuclear_2006}, energy~\cite{perkins_investigation_2000}, security~\cite{sowerby_recent_2007}, and medical science~\cite{wittig_boron_2008}.
One mechanism to generate ultra-short burst of fast neutrons is by interactions of laser-produced light ion beams with solid targets~\cite{roth-bright-2013, davis_neutron_2010, yang_neutron_2004, jung_characterization_2013, lancaster_characterization_2004,higginson_laser_2010, storm_fast_2013, mirfayzi_calibration_2015,kar_beamed_2016}.
Neutrons can also be generated through nuclear fusions in laser-produced plasmas~\cite{ditmire_nuclear_1999, bang_experimental_2013, zhao_novel_2016} or $(\gamma,\mathrm n)$ reactions~\cite{Photonuclear-PRL2000, pomerantz-ultrashort-2014}. 
Diagnosing pulsed neutrons is not only prerequisite to optimization of these neutron sources for applications, but also an essential tool for understanding fundamental physics in plasmas~\cite{alvarez_laser_2014, bang_calibration_2012}.
For example, in inertial confinement fusion (ICF) studies, plasma temperature and density distribution could be retrieved through the measurement of neutron yield and energy spectrum~\cite{cable_neutron_1987, kodama_fast_2001}.

Several techniques have been adopted to diagnose fast pulsed neutrons, including neutron active analysis~\cite{landoas_absolute_2011}, 
neutron track detector~\cite{oda_application_1991}, 
thermal neutron gas counter~\cite{moreno_system_2008}, along with neutron TOF~\cite{glebov_national_2010}, etc.
In neutron active analysis, foils made from silver, indium, or copper are commonly employed for absolute neutron dosimetry~\cite{slaughter_highly_1979}. 
However, due to its low detection efficiency, this method requires a minimum neutron yield of $10^5$ per burst~\cite{tarifeno-saldivia_calibration_2014}. 
The same problem is encountered by nuclear track detectors like CR-39~\cite{frenje_absolute_2002, kar_beamed_2016, bang_calibration_2012}.
Gas neutron counters (filled with $^3\mathrm{He}$ and $\mathrm{BF}_3$ etc.) 
usually employ external moderators to convert fast neutrons to slow neutrons.
The neutron related signals are spread in a time span of hundreds of microseconds as the fast neutron slowing down.
Due to gas detector's large pulse width (microseconds) and long dead time (tens of microseconds), it could have serious pile-up problem at high counting rate cases.

TOF technique plays a major role in neutron spectroscopy~\cite{vlad-time-1984}.
A TOF system usually consists of scintillators and fast photomultiplier tubes (PMTs), and could achieve high efficiency, as well as a good temporal response. 
Two serious issues emerge when TOF technique is employed in ultra-intense laser-plasma interactions, especially for solid targets.
Firstly, laser-induced electromagnetic pulses (EMP) and gamma-rays
may saturate the PMT and blind the detector for over hundreds of nanoseconds, 
making it difficult to recover neutron TOF signals within this temporal window.
To reduce the original gamma shower, lead shieldings are prerequisite to cover the scintillator detectors~\cite{roth-bright-2013}, which inevitably decreases the sensitivity.
Most neutrons generated from laser-induced nuclear reactions are fast neutrons, which have an energy of hundreds of keV or higher.
These fast neutrons arrive at detectors in hundreds of nanoseconds, when the PMT may not restore its sensitivity from the initial gamma-flash. 
Secondly, the pile-up effect caused by multiple neutrons at nearly the same time might drain the PMT voltage supply even though the detectors are operated in current mode~\cite{meigo_development_2000}. 

In this paper, we report a long-temporal-window TOF measurement.
Two temporal structures are identified and correlated to the neutron production, which are fast neutrons and delayed neutron-capture gamma rays.
We found a linear correlation between the fast neutron yield and the gamma-ray number, which provide a new way to diagnose laser-produced fast neutrons.
The basic idea is similar to the gas counters in neutron moderators, while the pulse width for a scintillator is much smaller, giving a superior higher temporal resolution and consequently a higher counting rate limitation.

\section{Experimental Setup}
\begin{figure}
 \centering
 \includegraphics[width=0.21\textwidth]{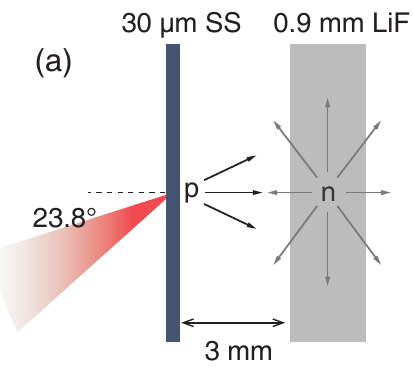}
 \hspace{0em}
 \includegraphics[width=0.26\textwidth]{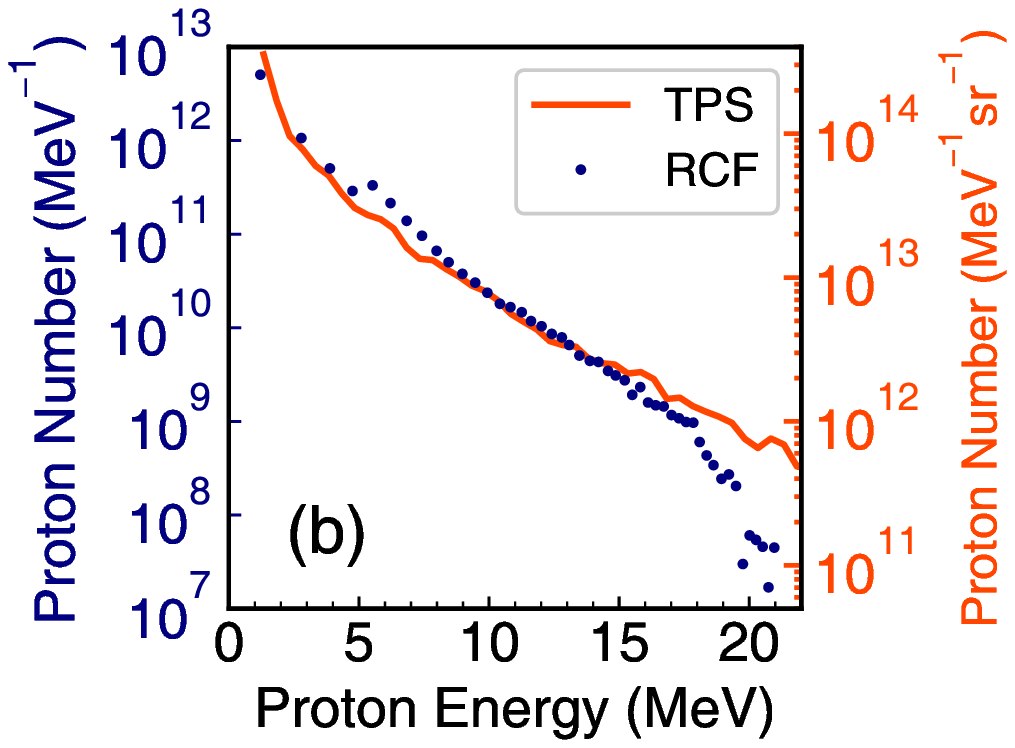} \\
 \hspace{-2em}
 \includegraphics[width=0.2\textwidth,trim=0 0 0 0]{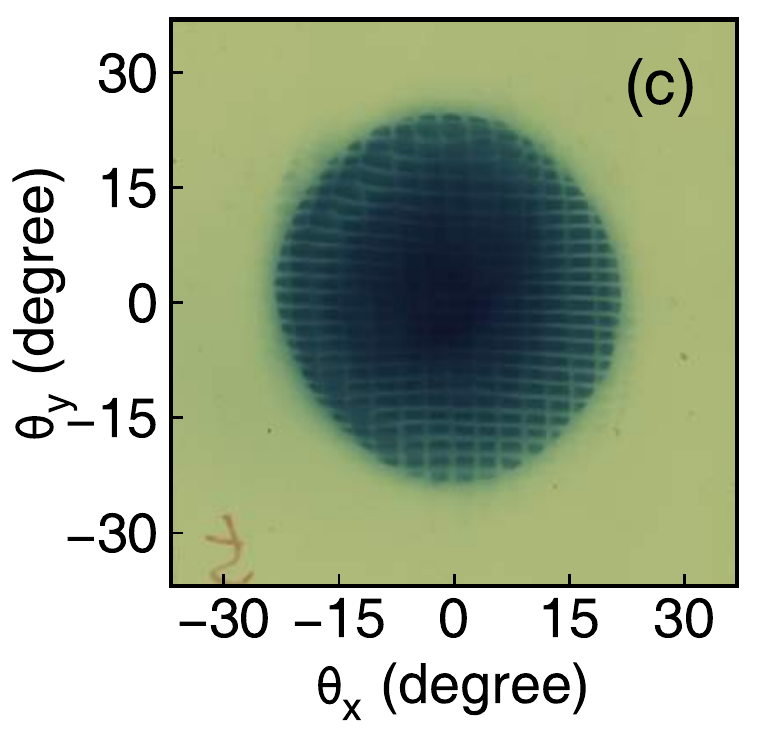}
 \hspace{0.5em}
 \includegraphics[width=0.2\textwidth,trim=0 0 0 0]{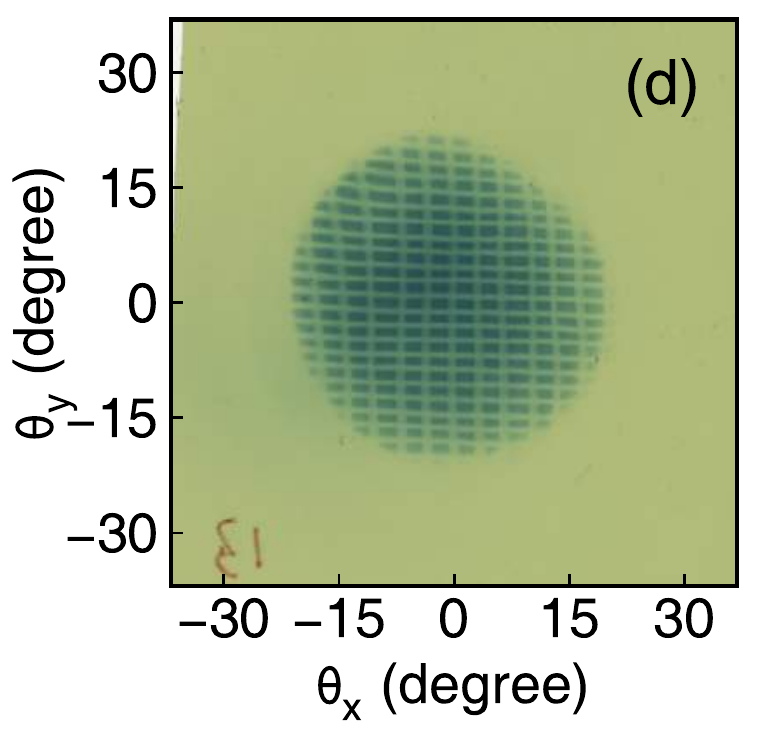}
 \caption{(a) Schematic of the experimental setup. The laser beam was focused on a 30um-thick stainless steel (SS) foil to generate protons by the TNSA mechanism. Accelerated protons hit on a 0.9 mm-thick LiF foil 3 mm apart and initiate the $\mathrm{^7Li(p,n)^7Be}$ reactions. 
 (b) Proton energy spectra measured from the Thomson parabola spectrometer (red line) and RCF stack (blue dots).
 (c)-(d) Selected RCF images showing the spatial distribution of 5.5 MeV and 10 MeV protons, respectively. The grid structures were shadows of a mesh inserted between proton reference target and the RCF stack.}
 \label{fig.setup}
\end{figure}

The experiment was performed utilizing the PW laser system at the Shenguang II (SG-II) facility, in the National Laboratory for High Power Lasers and Physics, Shanghai, China. 
The laser wavelength was 1053 nm and pulse duration was 0.7 ps at full width at half maximum (FWHM). The beam was focused onto the target with a spot size of 70 $\mu$m at an angle of $23.8^\circ$. The energy on target was 150 J in this experiment, 
giving rise to a peak intensity of 2.6$\times 10^{18}$ W/cm$^2$. 
A pitcher-catcher scenario of target was adopted for neutron generation.
The pitcher is a stainless steel (SS) foil with a thickness of 30 $\mu$m,
and the catcher is a 0.9 mm-thick LiF with the transverse dimensions of $3\times3\mathrm{mm}^2$, located at 3 mm behind the pitcher. 
The schematic is shown in Fig.~\ref{fig.setup}(a). 
A single-layer 30 $\mu$m-thick SS was also employed for reference.

The spatial-intensity distributions of the accelerated protons were measured by stacks of radiochromic film (RCF). The energy spectra of the sampled beam were measured by a Thomson parabola spectrometer (TPS) with an acceptance angle of $1.57\times 10^{-6}$ sr.
Calibrated image plate (IP) was used as detector in the TPS. 

Two types of TOF equipped with different scintillators were used for neutron detection. The first type is EJ-301 liquid scintillator with dimensions of $12.5\times \pi \times (12.5/2)^2$ cm$^3$. 
The EJ-301 has excellent pulse shape discrimination (PSD) properties which allow to discriminate the neutrons from gamma-ray.  
Six of these scintillators were placed around the target along different directions and shielded with lead brick houses of either 10 cm- or 5 cm-thick wall, respectively.
The second type is two BC-420 plastic scintillators ($10\times10\times40$ cm$^3$), which were also shielded in 5 cm-thick lead. The decay time of BC-420 is 1.5 ns, which is reduced by a factor of two in comparison to EJ-301, giving rise to a higher temporal resolution.
All the scintillators were coupled with fast photomultiplier tubes (PMTs), 
and the signals were recorded with oscilloscopes of 1 GHz bandwidth.

\section{Experimental Results and Discussion}

The energy spectra and spatial-intensity distributions of protons from the reference target were measured.
In Fig.~\ref{fig.setup}(b), the energy spectrum of protons sampled in a small solid angle by TPS is shown (red line), along with that of integrated whole beam recorded by the RCF stack (blue dots).
Both of them have near exponential distributions and the maximum proton energy is about 21 MeV.
Figure~\ref{fig.setup}(c)-(d) show two example spatial-intensity distributions of protons from the $\mathrm{5^{th}}$ and $\mathrm{13^{th}}$ RCF layers, corresponding to proton energies of 5.5 MeV and 10 MeV, respectively.
Well collimated beams are along the target rear normal direction. 
The beams have very high quality as seen from the shadowgraph of the crossed meshes.
Further data analyses show that the transverse emittance of the beams is less than 0.1 $\pi\, \mathrm{mm}\cdot\mathrm{mrad}$ and the virtual source size is less than 8 $\mu\mathrm m$.
In combination with the spectral shape, it suggests the dominant proton acceleration mechanism is the so-called target normal sheath acceleration (TNSA)~\cite{wilks-energetic-2001}.
These protons then impinge onto the LiF target and induce nuclear reactions 
$\mathrm{p + {}^7Li \rightarrow {}^7Be + n}$.

It has an energy threshold of 1.88 MeV and a notable cross-section up to 8 MeV.
The cross-section of the reaction 
$\mathrm{p + {}^6Li \rightarrow {}^6Be + n}$
for $^6\mathrm{Li}$ ($7.5\%$ natural abundance), another stable isotope of lithium, is too low to make contribution in neutron generation.

\begin{figure}
 \centering
 \includegraphics[width=0.48\textwidth]{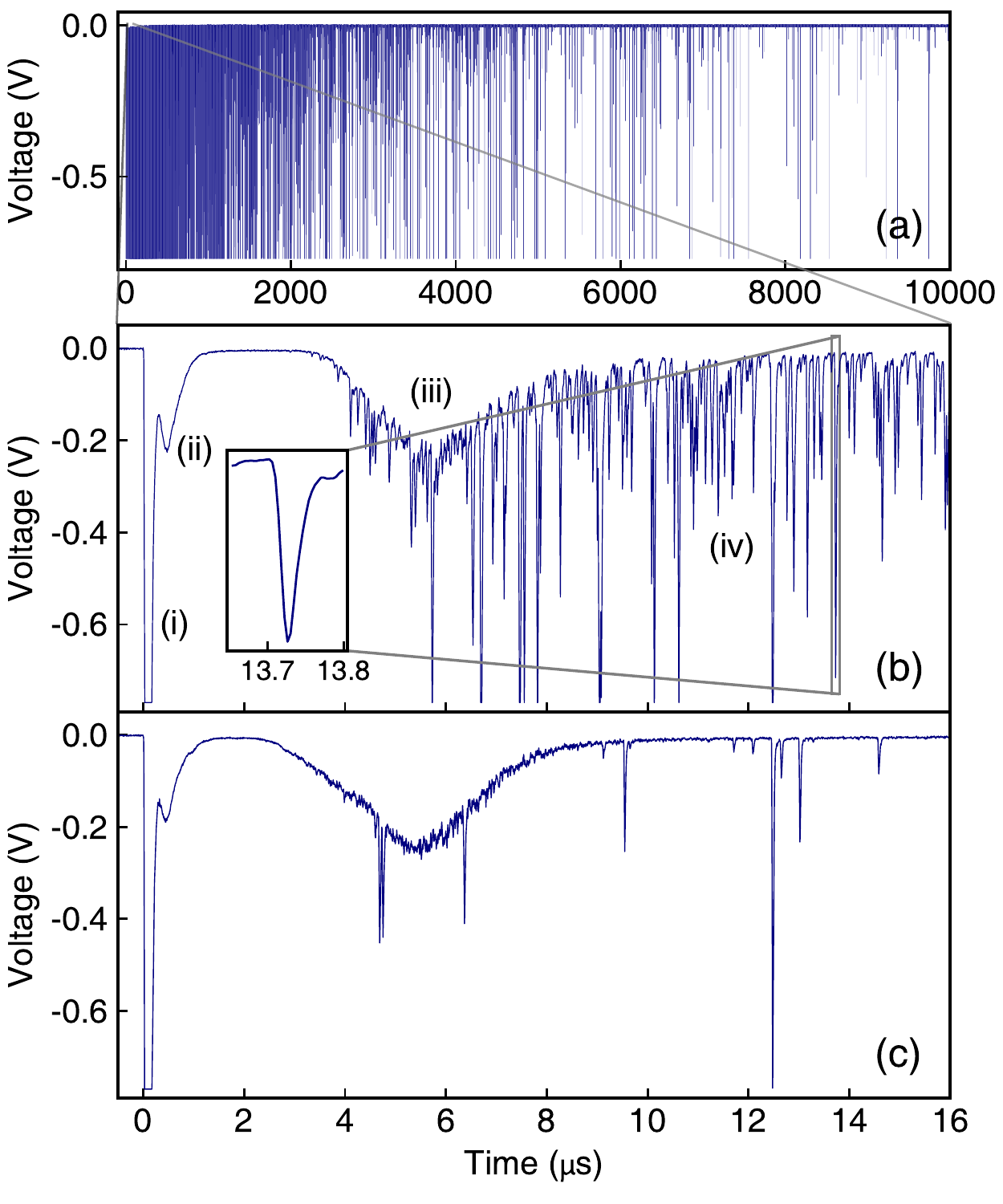}
 \caption{Example TOF signals. 
 (a) Overview in a time window of $0-10$ ms. Note the intense peaks within 1 ms.
 (b) Detailed view in a time window of $0-16$ $\mu\mathrm s$. 
 The inset of Fig.~\ref{fig.tof}(b) is the zoom-in of one single peak, which has a half-width of several tens nanoseconds. 
 (c) Waveform from a reference shot without a LiF catcher.
 }
 \label{fig.tof}
\end{figure}

Despite the differences in location and high-voltage setting for all the detectors, similar signal patterns were found.
A good repeatability was also measured from shot to shot for similar laser and target parameters.
A typical TOF signal recorded by one of the scintillators is shown in Fig.~\ref{fig.tof}.
Figure~\ref{fig.tof}(a) represents the overall signal ranging from 0 to 10 ms,
while Fig.~\ref{fig.tof}(b) is a section of Fig.~\ref{fig.tof}(a) ranging from 0 to 16 $\mu$s.

It can be seen that the temporal waveform has four characteristic structures, labeled from (i) to (iv) in Fig.~\ref{fig.tof}(b):
\begin{enumerate}[(i)]
\item \label{itm:first} The signal dip at $t=20$ ns with a width of 0.1 $\mu$s.
It accounts for the copious bremsstrahlung photons generated at the time $t=0$, 
i.e. the moment when the laser hits the target. 
Because the prompt radiation is too intense, the signal overrun the input voltage range of the oscilloscope.
This dip could be reduced by the extensive lead shielding;

\item \label{itm:second} The signal dip at $t=0.3$ $\mu$s with a width of 0.2 $\mu$s. This is the signal of fast neutrons with energies of $1-5$ MeV, which are produced from the $\mathrm{^7Li(p,n)^7Be}$ reactions;

\item \label{itm:third} The broad dip from 4 $\mu$ to 8 $\mu$s. This is overlapped with the leading part of structure (\ref{itm:fourth}) which is attributed to the malfunction of the PMT. 
Due to the strong gamma photon shower at $t=20$ ns, the current in the PMT circuit (especially between the last dynode and the anode) is very high, which makes the voltage between electrodes drops~\cite{PMT-2002-flyckt}.
This broad dip structure is caused by the voltage recovery, and the dip width is determined by the electrode restore time;

\item \label{itm:fourth} The intense narrow discrete peaks from approximately 4 $\mu$s to 10 ms. The typical width is 10 ns.
One zoom-in example of these peaks is shown in the inset of Fig.~\ref{fig.tof}(b).
This type of signal has not been reported as far as we know. In the following, we will focus on it.
\end{enumerate}

Signals of two PMTs coupled to a same plastic scintillator are compared.
The results show that the timing and amplitude of the peaks from the two PMTs are well matched.
This indicates that their signals are caused by incoming radiation events instead of false signals due to thermal noises of the PMT circuits.

When the reference target was used, i.e. shot without a LiF catcher, there remained a much smaller neutron background originated from $(\gamma,\mathrm n)$ reactions. The amplitude of structure (\ref{itm:second}) reduces, and peaks in structure (\ref{itm:fourth}) are scarce, while structure (\ref{itm:first}) and (\ref{itm:third}) almost remain the same as shown in Fig.~\ref{fig.tof}(c).
This suggests that structure (\ref{itm:second}) and (\ref{itm:fourth}) are correlated and (\ref{itm:fourth}) is due to neutron generation.

\begin{figure}
 \centering
 \includegraphics[width=0.48\textwidth]{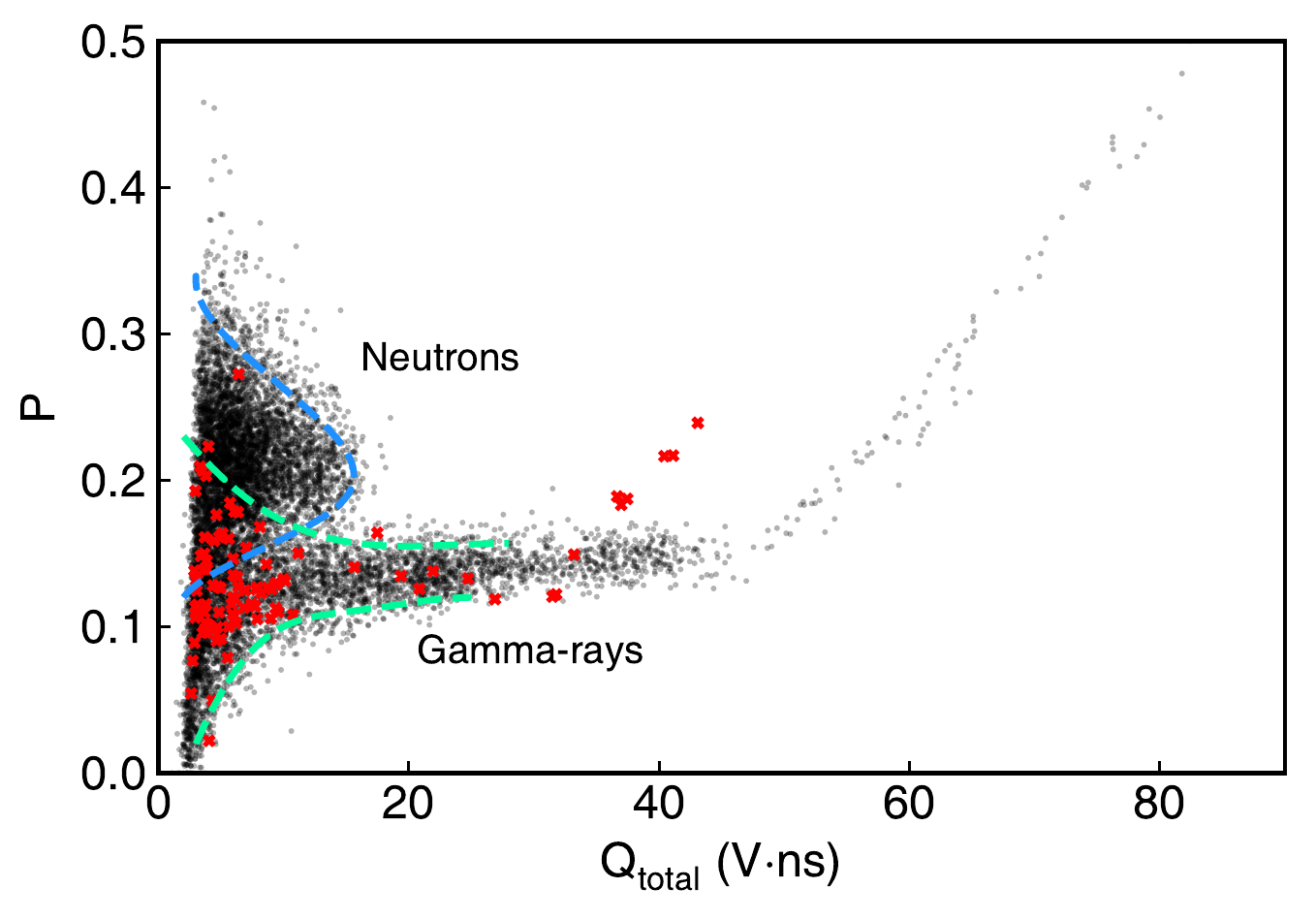}
 \caption{Pulse shape discrimination results. Pulses from a $^{252}\mathrm{Cf}$ source (black dots) are separated into two groups representing neutrons and gamma-rays respectively. Nearly all pulses from the experiment (red crosses) fall in the gamma region. The dashed lines are drawn to guide the eye.
  }
 \label{fig.psd}
\end{figure}

A pulse shape discrimination (PSD) procedure was conducted to distinguish between gamma-rays and neutrons by analyzing the difference in their characteristic pulse shapes~\cite{brooks_scintillation_1959}.
Pulses in structure (\ref{itm:fourth}) were analyzed by defining a PSD parameter $P={Q_{\mathrm{tail}}}/{Q_{\mathrm{total}}}$, where $Q_{\mathrm{total}}$ and $Q_{\mathrm{tail}}$ are the charge integration over the whole pulse and the tail part only, respectively~\cite{tomanin_characterization_2014}.
The results ($P$ versus $Q_{\mathrm{total}}$) are shown in Fig.~\ref{fig.psd} as red crosses.
To determine the respective parameter regimes for neutrons and gamma-rays, the detector was calibrated with a $^{252}\mathrm{Cf}$ radiation source, with a same high-voltage setting as in the experiment. 
$^{252}\mathrm{Cf}$ is a neutron emitter with strong gamma background, so that both neutrons and gamma-rays can be recorded and analyzed. 
The results are shown in black dots in Fig.~\ref{fig.psd}, in which one can see neutrons and gamma-rays are well distinguished with only a minor portion of overlap.
It is clear that most of the pulses from structure (\ref{itm:fourth}) are distributed in the gamma region.
Therefore, we conclude that these pulses are predominately attributed to gamma-rays.

\begin{figure}
 \centering
 \includegraphics[width=0.48\textwidth]{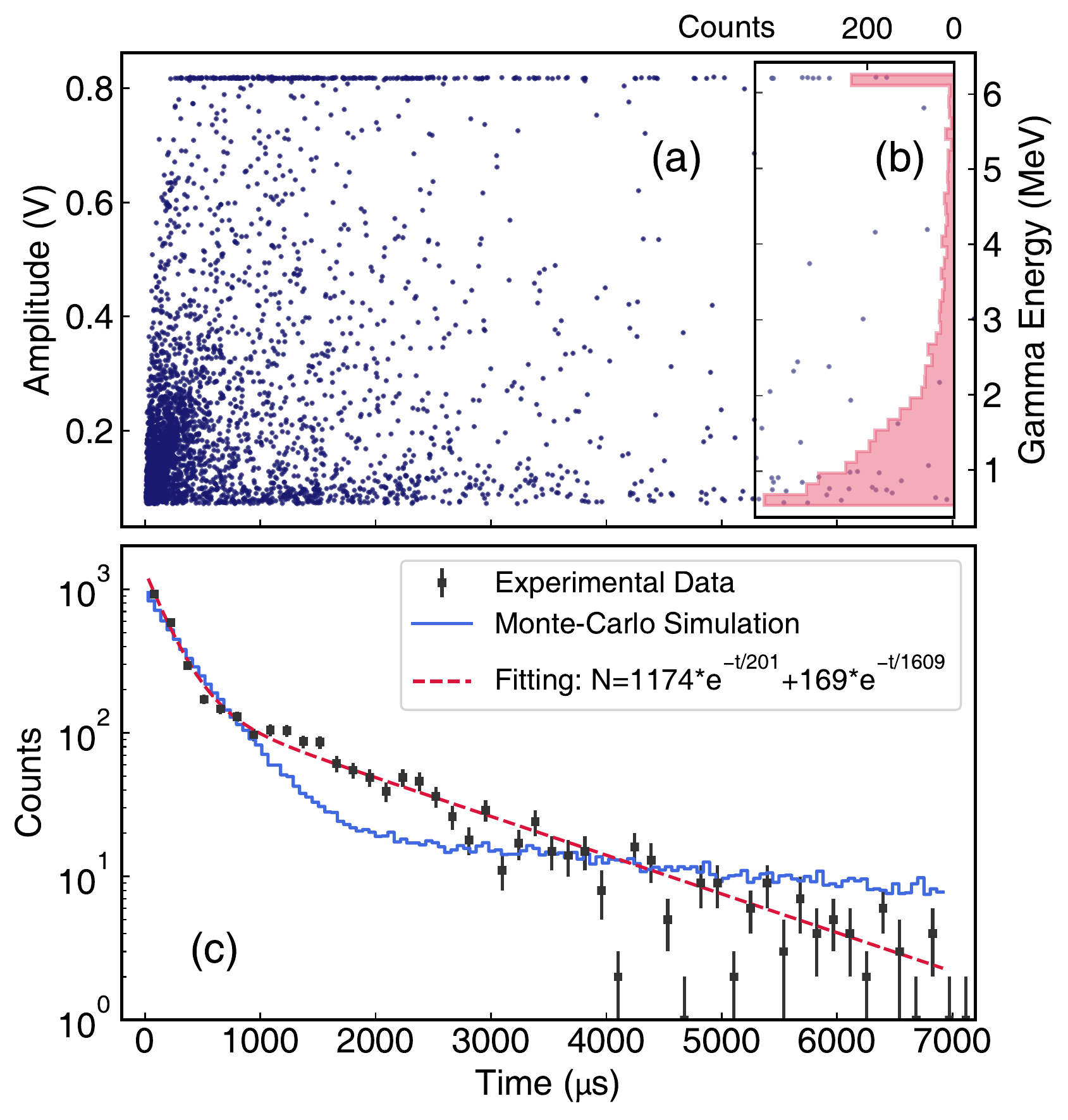}
 \caption{
 Temporal-amplitude distribution of the gamma-ray signals.
 (a) Scatter plot of gamma-ray events.
 (b) Amplitude projection of the scatter plot. The maximum value ($\sim 0.82$ V) is limited by the dynamic range of the oscilloscope, corresponding to a gamma-ray energy of 6.2 MeV.
 (c) Temporal projection of the scatter plot.
 The experimental data is fitted by a two-component exponential decay function 
 and shown as a red dashed line.
 The Monte-Carlo simulation is shown as a blue solid line.
  }
 \label{fig.tof.2d}
\end{figure}

A $^{137}$Cs and a $^{60}$Co gamma sources were further used to calibrate the detector, enabling the pulse amplitude to be converted into gamma-ray energy, by determining their positions of Compton edges in the response functions~\cite{tomanin_characterization_2014}. 
A peak-seeking algorithm was carried out to reconstruct the temporal and amplitude distributions of these delayed gamma-rays.
The result is shown in Fig.~\ref{fig.tof.2d}(a). 
The amplitude distribution is showin Fig. ~\ref{fig.tof.2d}(b). 
Gamma-ray signals with an amplitude greater than 0.82 V pile up at the maximum (corresponding to a gamma-ray energy of 6.2 MeV) due to the dynamic range of the oscilloscope.
The temporal spectrum (Fig.~\ref{fig.tof.2d}(c)) shows a double exponential decay feature, and can be fitted by function (red dashed line):
\begin{equation}
N=(1174\pm32)\cdot e^{-t/(201\pm11)}+(169\pm23)\cdot e^{-t/(1609\pm192)}.
\end{equation}

It is theoretically possible that these gamma-rays are from radioactive isotopes or isomers directly created by the high-intensity laser.
However, after searching the chart of nuclides for a candidate, 
we could not find a known isotope with a half-life of several tens of microseconds that fits in our experimental conditions.

On the other hand, fast neutrons generated by $\mathrm{{}^7Li(p,n){}^7Be}$ reactions interact with the target chamber, air, as well as with the supporting infrastructures outside the chamber. 
After tens of scattering events, a neutron reaches thermal equilibrium with the atoms of the medium. 
As the neutron energy decreases, its capture cross-section increases.
Considering that the life of a free neutron is approximately 10 minutes, almost all neutrons end up with being captured. 
The compound nucleus that has absorbed a neutron subsequently decays to its ground state with prompt emission of one or more characteristic gamma-rays~\cite{foderaro_elements_1971}.
The process could be formulated as 
${}^A\mathrm{X}_Z + \mathrm{n} \rightarrow {}^{A+1}\mathrm{X}^{\ast}_Z \rightarrow {}^{A+1}\mathrm{X}_Z + \gamma$.

Neutron capture cross-sections for some isotopes existed in our experimental environment are listed in Table.~\ref{tab.n.CS}.
One of them 
$\mathrm{{}^{56}Fe+n \rightarrow {}^{57}Fe^{\ast}}$
has a fairly large capture cross-section. 
A considerable portion of the gamma-rays have energies of $6-8$ MeV from the gamma spectrum of ${}^{57}\mathrm{Fe}^{\ast}$ decay (evaluated nuclear database~\cite{reedy_prompt_2002}). 
This is in agreement with our observation in Fig.~\ref{fig.tof.2d}(b).
The reaction of 
$\mathrm{{}^{1}H+n \rightarrow {}^{2}H^{\ast}}$
is another candidate where hydrogen also has a large neutron capture cross-section, 
and is widely distributed in the concrete infrastructures of the experimental hall.
However, it makes no contribution to the higher gamma-ray energies,
since the only gamma-ray line emitted by hydrogen neutron capture has an energy of 2.2 MeV~\cite{reedy_prompt_2002}.

\begin{table}
\caption{\label{tab.n.CS}Thermal neutron absorption cross sections $\sigma_n$ 
and corresponding maximum gamma-ray energy $E_\gamma^{\mathrm{max}}$ 
for isotopes which were in relatively large abundances in our experimental environment. 
Taken from Ref.~\cite{reedy_prompt_2002}.}
\begin{tabular}{r d d}
\toprule
Isotope       &\multicolumn{1}{c}{$\sigma_n\ (\mathrm b)$}  &\multicolumn{1}{c}{$E_\gamma^{\mathrm{max}}\ \mathrm{(MeV)}$} \\
\colrule
$^1$H         & 0.332         & 2.22\\
$^{12}$C      & 0.0035        & 4.95\\
$^{14}$N      & 0.080         & 10.83\\
$^{16}$O      & 0.00019       & 4.14\\
$^{27}$Al     & 0.230         & 7.73\\
$^{28}$Si     & 0.17          & 8.47\\
$^{56}$Fe     & 2.6           & 7.65\\
$^{63}$Cu     & 4.5           & 7.92\\
\botrule

\end{tabular} 
\end{table}

In the case of a point neutron source in an infinite moderator,
the temporal profile of the neutron captures can be described by exponential functions~\cite{bowden_note_2012}, 
whose slopes depend on the material and initial energy of the neutrons. 
In a real experiment, the geometry and materials can be very complicated.
Therefore, we take a Monte-Carlo (MC) toolkit Geant4~\cite{agostinelli_geant4simulation_2003} 
to simulate the neutron transport and moderation under our experimental conditions.
The geometry for simulation includes the main structure of the target chamber with an isotropic point neutron source at its center, diagnostic apparatus, along with large objects inside and outside the chamber, ceiling, floor, and walls of the experimental hall. 
Physics processes involved in neutron-matter interactions including elastic and inelastic scattering, neutron capture, are considered.
The simulated temporal distribution of neutron captures (Fig.~\ref{fig.tof.2d}(c)) agrees with the first parameter of the experimental data while it overestimates the second.
We suspect that the disagreement may be a result of the simplification in MC geometry modeling.

To quantitively investigate the correlations of the number of radiative capture gamma-rays and the fast neutron yield,
we conducted another neutron-generation experiment.
In this particular experiment, two sets of laser beams were used, 
each delivering 4$\times$250 J energy onto two targets separated by 4.4 mm.
Each target has a $0.5\times 0.5\ \mathrm{mm}^2$ copper base and was coated with 10 $\mu$m thick deuterated hydrocarbon (CD) layer.
Each laser had a typical pulse width of 1 ns and a focal spot diameter of 150 $\mu$m. 
Ablated plasmas expand and collide with each other in the middle area between the targets, 
inducing $\mathrm{D(d,n){}^3He}$ reactions, and generating monoenergetic neutrons with an energy of 2.4 MeV. 
Similar liquid and plastic scintillation detectors were used for neutron detection.
A detailed description of the experiment can be found in Ref.~\cite{zhao_neutron_2015}.
Comparing with picosecond-duration laser driver, backgrounds caused by bremsstrahlung gamma-ray bursts and EMP noise in nanosecond pulse lasers are much smaller.
Thus the neutron yields can be determined more accurately.

\begin{figure}
 \centering
 \includegraphics[width=0.45\textwidth]{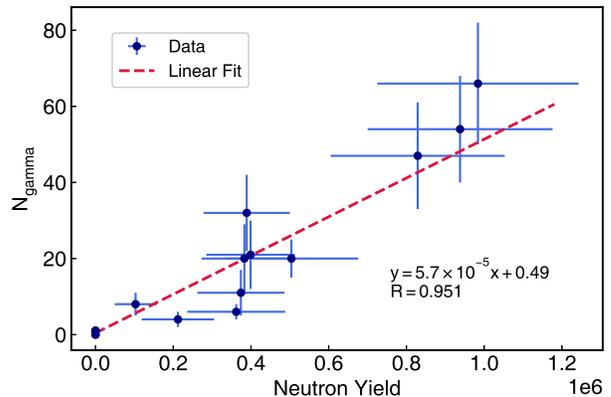}

 \caption{
 Observed gamma-ray number versus neutron yield. The dashed line is a linear fit to the experimental data.
 }
 \label{fig.n-gamma-ratio}
\end{figure}

The dependence of the number of delayed gamma-rays $N_{\gamma}$ and the neutron yield $N_{\mathrm n}$ is shown in Fig.~\ref{fig.n-gamma-ratio}. 
One can find a linear relationship between $N_\gamma$ and $N_{\mathrm n}$.
The Linear fitting correlation coefficient is $R=0.951$.
A slight underestimation on $N_\gamma$ for the small neutron yield was observed.
This may be due to the insufficient statistic of the gamma-ray numbers and the large uncertainty of the neutron yields.
With these results, we conclude that $N_\gamma$ can be seen as a prompt and convenient parameter for the estimation of the fast neutron yield.

\section{Summary}
In summary, delayed discrete gamma-rays with energies of MeV level were observed in a time window from several microseconds to milliseconds using a TOF diagnostic. 
The number of gamma-rays has a linear correlation with the fast neutron yield on a shot-by-shot basis.
A diagnostic of pulsed fast neutrons can benefit from measuring the thermal neutron related gamma-rays with commonly used TOF systems. 
Fast neutron signals from a scintillation detector in high-intensity laser experiments are often obscured by bremsstrahlung X-rays and EMP interference, 
while delayed (n,~$\gamma$) signals provides a quick reference on whether or how many neutrons are generated. 
Because these gamma signals are separated temporally with the initial X-ray and EMP bursts.
High sensitivity and signal-to-noise ratio is expected by detecting both fast and thermal neutron related signals with the same detector.

This method can also be used in other fundamental physics studies under extreme conditions which cannot be provided by conventional accelerators.
For instance, in a hot and dense plasma produced by high-intensity lasers, 
new nuclear excitation states, isomers, and other exotic atomic states can be created 
but difficult to detect~\cite{savelev-direct-2017, andreev_excitation_2000}.
The method shown here can detect those states with lifetimes in the range from about 100 ns to 1 s.

\section{Acknowledgements}

We acknowledge financial supports from 985-III grant from Shanghai Jiao Tong University, National Basic Research Program of China (Grant Nos. 2013CBA01502 and 2010CB833005), National Natural Science Foundation of China (Grant Nos. 11375114 and 11205100), Doctoral Fund of Ministry of Education of China (Grant No. 20120073110065), Strategic Priority Research Program of the Chinese Academy of Sciences (Grant Nos. XDB16010200 and XDB07030300), and Shanghai Municipal Science and Technology Commission (Grant No. 11DZ2260700). 
The authors thank the staff of the SG-II laser facility for operating the laser and target area.


%

\end{document}